# Artificial Neural Network for LiDAL Systems

Aubida A. Al-Hameed[1], Safwan Hafeedh Younus[1], Ahmed Taha Hussein, Mohammed T. Alresheedi[2] and Jaafar M. H. Elmirghani[1]
[1]School of Electronic and Electrical Engineering, University of Leeds, LS2 9JT, United Kingdom
[2]Department of Electrical Engineering, King Saud University, Riyadh, Saudi Arabia
{elaawj, elshy @leeds.ac.uk, asdftaha@yahoo.com , malresheedi@ksu.edu.sa, j.m.h.elmirghani@leeds.ac.uk}

*Abstract*—In this paper, we introduce an intelligent light detection and localization (LiDAL) system that uses artificial neural networks (ANN). The LiDAL systems of interest are MIMO LiDAL and MISO IMG LiDAL systems. A trained ANN with the LiDAL system of interest is used to distinguish a human (target) from the background obstacles (furniture) in a realistic indoor environment. In the LiDAL systems, the received reflected signals in the time domain have different patterns corresponding to the number of targets and their locations in an indoor environment. The indoor environment with background obstacles (furniture) appears as a set of patterns in the time domain when the transmitted optical signals are reflected from objects in LiDAL systems. Hence, a trained neural network that has the ability to classify and recognize the received signal patterns can distinguish the targets from the background obstacles in a realistic environment. The LiDAL systems with ANN are evaluated in a realistic indoor environment through computer simulation.

## I. Introduction

Visible light communication (VLC) is part of optical wireless communication (OWC) that uses light as a carrier to modulate the information signal in the visible spectrum (380nm to 780nm) [1-4]. VLC systems are becoming more popular everyday due to their inherent advantages over radio frequency (RF) systems. The advantages include a large unregulated spectrum, low complexity of transceiver units, freedom from fading, confidentiality and immunity against interference from electrical devices [5]-[8]. VLC system applications can support indoor high data rate communication [6], [8] under-water communication [9], [10], LED to LED communication [1], [11] and indoor user localisation [12]-[13]. In [14], a light sensing system using VLC (LiSense) was proposed to track the human gesture and reconstruct human skeleton. The LiSense system makes use of 324 photodetector array placed on the floor to sense the beacon signals sent from the light sources (VLC transmitters) to recover the human shadow pattern created by individual VLC transmitters. A laser radar in conjunction with VLC system was introduced in [15] to provide vehicle to vehicle ranging and VLC communication.

People detection and counting in an indoor environment (such as in offices, exhibition halls and shopping malls) can provide useful information for different applications [16]-[18]. For example, human presence detection is valuable for security purposes. Also knowing the number of people in a supermarket may have an important practical use in terms of marketing, management, optimisation and maintaining a high quality of service.

Human sensing (i.e. human detection and counting) in an indoor environment is a very challenging endeavour for many reasons; (*i*) sudden changes may occur in the environment conditions. For instance in an outdoor environment, RADAR signals can be affected by the rain or fog while for an indoor environment passive infraed (PIR) sensors can be activated wrongly by heat currents from heating and air conditioning [19]-[21]; (*ii*) the reflected signal from the background is very similar to the one reflected by a person, thus separating a person from the background is an essential requirement for human sensing in a realistic environment. Also, for RADAR and LADAR sensing systems, the reflected received signal suffers from multipath propagation leading to fooling the sensing system and to false person detection (phantom detection) [19]; and (*iii*) people behaviour is unpredictable with a high degree of similarity such as walking in random paths that may change suddenly resulting in a serious challenge to localise and track individuals correctly [19].

Ultra-wideband (UWB) RADAR systems with a transmitted signal bandwidth greater than 500 MHz have been introduced to detect, localise and track humans in the indoor environment [20], [22], [23]. The UWB carrier signal with a typical frequency range of 3.1GHz to 5.3 GHz can penetrate walls, furniture and human body [24]. This enables UWB RADAR systems to support various applications such as human movement detection through-walls for security applications and biomedical applications (i.e. monitoring human vital signs) [25], [26]. In UWB RADAR, detection of the target (human) depends on the target motion where the human movement causes changes in frequency, phase and time of arrival. However for UWB radar employed in an indoor environment, the effects of signal scattering and absorption by obstacles significantly impair the performance of UWB indoor radar [19], [20].

Binary sensors such as Passive Infrared (PIR) sensors, break beam and binary Doppler sensors have been used to detect human presence and rely on the human motion [19], [27], [28]. The main drawback of binary sensors is their large false detection. For example, the PIR system is temperature dependent, thus any change in the environment temperature leads to a vast number of detection failures. Doppler shift sensors use the concept that signals reflected from a mobile object suffer a frequency shift depending on the object's speed. The Doppler shift sensor can provide a speed measurement of the detected human unlike the PIR sensor. In [29] a one dimensional Doppler radar was proposed to detect stationary humans relying on the motion of human breathing lungs. A laser radar (LADAR) has been used to detect people based on their shape through extracting high resolution two and/or three dimensional snapshots of the environment [30], [31]. In [32] a single 360-degree LADAR system was introduced to detect and track people in an indoor environment. However, the main disadvantage is the system complexity, eye safety due to the laser beam and the relatively long time needed to scan the environment with high resolution which may lead to miss detecting humans walking at a fast pace.

A light detection and localization (LiDAL) system was proposed in [33] for detection, counting and localisation in an indoor setting. The LiDAL system focuses on human sensing to provide people with spatio-temporal indoor localization information. It carries out presence detection, counting, localization and tracking. In this application, people can be distinguished from the background due to their dynamic characteristics that arise from their activity (siting/standing) and motion (walking), while stationary people are undetectable [33].

In this paper, we expand the work proposed in [33] for people detection, counting and localization using LiDAL systems namely; MIMO LiDAL and MISO IMG LiDAL systems. We introduce an intelligent LiDAL system that uses artificial neural networks (ANN). Based on our observations in the LiDAL systems introduced in [33], the received reflected signals in the time domain have different patterns corresponding to the number of targets and their locations in an indoor environment. The indoor environment with obstacles (furniture) appears as a set of patterns in the time domain when the transmitted optical signals are reflected from objects in MIMO LiDAL systems. The patterns appear in the spatial domain in the imaging receiver pixels in MISO IMG LIDAL systems. When targets enter the environment, they add to / change the temporal and spatial reflection patterns in the room. Therefore, a trained neural network that has the ability to classify and recognize the received signal patterns can distinguish the targets from the background obstacles in a realistic environment.

This paper is divided into section as follows: Section II considers the design of the MIMO LiDAL and MISO LiDAL systems. Section III introduces the ANN training used in the LiDAL systems. Section IV presents the simulation setup, the target mobility model and realistic indoor environment. Section V presents the results and discussion. Finally, conclusions are drawn in Section VI.

## II. LiDAL Systems

In this section, we introduce the configuration of the MIMO LiDAL system and the MISO IMG LiDAL system which were proposed in [33]. In addition, we present distinguishing methods namely; background subtraction method and cross correlation for mobile target distinguishing from the background obstacles (furniture).

### A. MIMO LiDAL System

The MIMO-LiDAL system is used to detect, count and localize targets [33]. The system consists of multiple LiDAL transmitters and multiple LiDAL receivers. The MIMO-LiDAL system employs a single photodetector receiver collocated with each VLC transmitter (luminaire, light source) which represents a transceiver unit ($T_{RX}$). The MIMO-LiDAL system has eight transceiver units placed in the room ceiling as can be seen in Fig.1. Each transceivers unit cover a circular optical detection zone with a radius of 1.25m. The transceiver units are spaced by 2m [33]. In the MIMO LiDAL system considered, the total number of optical detection zones was 8. The MIMO-LiDAL system is designed to resolve the ambiguity of target detection and localization by implementing collaboration between the neighbouring transceiver units [33]. The target localization is tackled by joint use of three transceiver units working together through three system scans (three consecutive listening (frame) times). The MIMO-LiDAL listening time is divide into $N$ time slots where the time slot width is 2ns (equal to the transmitted pulse width) which enables a ,$\Delta R$, 30cm target detection resolution for MIMO LiDAL system as reported in [33]. A time of arrival (TOA) technique is used to calculate the target range from a transceiver unit [33]. The target range calculation is based on the trip time of the reflected pulse from the target and the speed of light. The three ranges obtained from the transceiver units are used with a triangulation method to determine the intersection of the (circles) ranges resulting in an estimated target location on the detection floor [33].

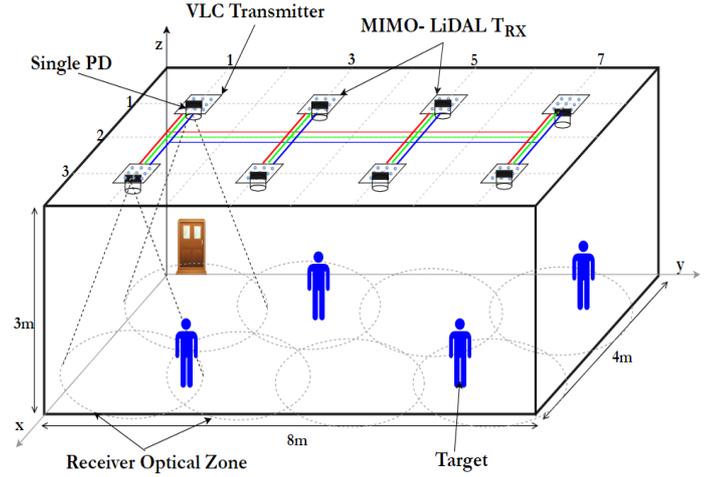

Fig. 1: MIMO LiDAL System.

### B. MISO IMG LiDAL System

The MISO IMG LiDAL system is used to detect, count and localize targets. The system employed multiple LiDAL transmitters units with a single imaging receiver. In the MISO IMG LiDAL system, the imaging receiver was placed in the centre of the room's ceiling as can be seen in Fig. 2 [33]. In MISO IMG LiDAL system design, the imaging receiver has an array of (8 columns × 16 rows) pixel receivers [33]. The MISO IMG LiDAL system forms an image of $N_P = 128$ pixels where every pixel receiver covers a narrow optical detection zone. The benefits of the massive number of pixels are in enabling spatial selection to separate the targets in multiple narrow optical zones. Thus the MISO IMG LiDAL system can: (i) eliminate the ambiguity of target detection and localization and (ii) minimize the interference resulting from the reflections of the background obstacles [33]. In the MISO IMG LiDAL system, the target detection resolution $\Delta S$ was 0.5m (i.e. the LiDAL system is able to separate two targets at a distance of 0.5m) [33]. The MISO IMG LiDAL system employs a direction of arrival (DOA) method to determine the target location where the image of the target in the pixels that cover the optical zones determines the target location [33]. It is worth mentioning that, the imaging receiver of 128 pixels was divided into 8 groups with 16 pixels per group [33]. Each group of receiver pixels (GRP), works separately with one transmitter during the MISO IMG LiDAL scan (snapshot).

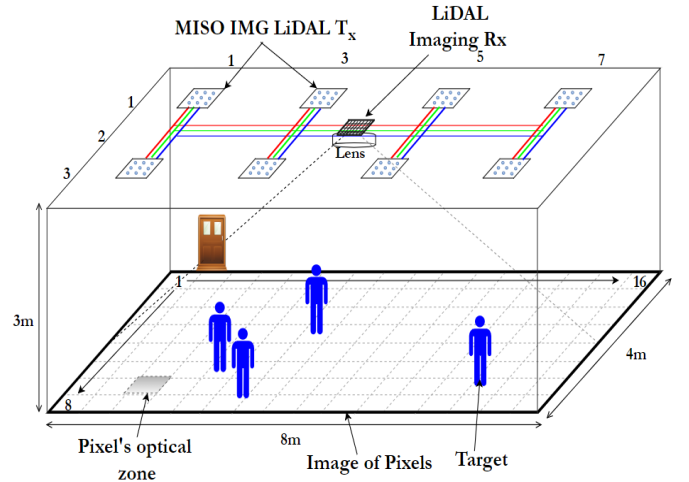

Fig. 2: MISO IMG LiDAL System.

## C. Background subtraction method (BSM)

The background subtraction method (BSM) was introduced and examined in [33]-[35]. In this method, the received reflected signals during multiple radar scans are subtracted in order to distinguish the moving target. The BSM relies on the fact that the received signal from the background obstacles (furniture) is stationary (time-invariant) and the received signal reflected from the target is time-variant due to its motion. It is worth mentioning that subtracting two or more received signals leads to enhancing the variance of the white Gaussian noise. Moreover, the performance of BSM is significantly degraded in case the mobile target moves in the horizontal distance where its signal reflections arrive at the same time during radar scans resulting in miss-distinguishing the mobile target [34]-[36].

## D. Cross Correlation method (CCM)

The cross-correlation method (CCM) was proposed and examined in [33]. In this method, the mobile target can be distinguished relative to the stationary background obstacles by correlating the received reflected signals through multiple LiDAL snapshot measurements in order to monitor the change due to target mobility. It is worth mentioning that the cross-correlation method has better performance than the Doppler method in terms of detecting the change of targets of interest when the targets move at low speed [37]. Also, the cross-correlation has the benefit of detecting weak received reflected signals and is more robust in the presence of noise [33], [38].

## III. NEURAL NETWORK FOR LIDAL SYSTEMS

In this section, we introduce the ANN with the MIMO LiDAL and MISO IMG LiDAL systems. Many algorithms have been employed to train artificial neural networks. The backpropagation (BP) algorithm is one of the most popular approaches that have been used to train neural networks due to its efficiency and simplicity [39], [40], [41]. In this paper, a supervised learning algorithm with backpropagation is deployed to train multi-layer neural networks.

In the MIMO LiDAL systems investigated in this paper, we employed an ANN in each light unit (transceiver) and therefore this ANN covers one optical zone. This ANN has to measure one of the three distances needed to localize a target. The other two distance measurements (bistatic or monostatic measurements) are then carried out by uploading different weights in the ANN.

The received reflected signals were sampled at a sampling rate of $1/T_{sa}^{MIMO}$, and the samples were grouped into $N_{slot}$ time slots. The number of input nodes in the ANN is equal to the total number of samples taken, with each group of samples (ie each group of ANN nodes) labelled as a time slot. The time slot duration is $T_s$ and therefore, the total number of input nodes in the ANN is $N_{in}^{MIMO} = N_{slot}(T_s/T_{sa}^{MIMO})$ as can be seen in Fig. 3. The output of the ANN has $N_{slot}$ nodes, where the output of each node is a one or zero indicating the presence or absence of a target in that time slot. Hence, the ANN is firstly trained in the given optical zone in the presence of furniture (obstacles) and the absence of targets, resulting in an all zero output. The ANN is then trained in the presence of single or multiple targets, with the supervision indicating the time slot that has a target (one).

For the MISO-IMG-LiDAL system, the number of ANNs used to cover the entire room is equal to $L_{tx}$, the number of active transmitters, ie the number of 'groups of pixels' (see Fig. 5) [33]. For example, in the room in Fig. 2, $L_{tx} = 8$, thus 8 ANNs are needed in the room. The total number of pixels in the imaging receiver is $N_P$, and therefore the number of pixels observed by each ANN is $N_P/L_{tx}$. A snapshot / image is taken by the imaging receiver every $T_{sa}^{IMG}$ which is our sampling period here. Note that the input to the ANN should have more than a single time sample per pixel to result in improved robustness against variations in the environment. In the MIMO LiDAL system, $T_s/T_{sa}^{MIMO}$ samples were taken per time slot to help the ANN deal with channel and environment (obstacles) impairments. The imaging receiver pixels see $\Delta S$=0.5m on the room floor, while MIMO LiDAL system has $\Delta R$ =0.3m on the floor [33]. Therefore, we increased the number of samples taken by the MISO-IMG-LiDAL system by a factor $\Delta S/\Delta R$. For fairness, we also set the sampling rate in the two systems to the same value, $T_{sa}^{IMG} = T_{sa}^{MIMO}$. Hence, the number of input nodes in the MISO-IMG-LiDAL ANN, $N_{in}^{IMG}$, is

$$N_{in}^{IMG} = \frac{T_s}{T_{sa}^{MIMO}} \frac{\Delta S}{\Delta R} \frac{N_P}{L_{tx}} \qquad (1)$$

The MISO-IMG-LiDAL ANN has $N_P/L_{tx}$ outputs where each output represents a pixel and indicates in a binary fashion the presence or absence of a target in the FOV of that pixel. Next we will introduce the methods used in training the ANNs.

### A. The neural network training process

The neural network consisted of an array of inputs (input layer), hidden neurons (hidden layer) and one output layer as shown in Fig. 3. The number of hidden neurons is important in determining the performance of the neural network. For a neural network consisting of a vast number of hidden neurons, the following observations hold: (*i*) it is possible to over-fit, (*ii*) the neural network complexity increases (iii) the ANN learns the exact training samples and this reduces its ability to recognize new signal patterns [41], [42]. On the other hand, a neural network with a few hidden neurons may have a limited learning memory with inadequate performance (under-fitting) [42]. In this paper, we considered a pruning approach in order to optimize the number of hidden neurons ($N_m$) [43]. The number of hidden neurons is calculated as [44]:

$$N_m = \frac{(N_{in}+Y)}{\beta_i} \qquad (2)$$

where $Y$ is the number of output neurons ($Y = N_P/L_{tx}$ or $Y = N_{slot}$ for the MISO-IMG-LiDAL and MIMO-LiDAL systems respectively), $N_{in}$ is the number of the input neurons (given by $N_{in}^{IMG}$ or $N_{in}^{MIMO}$ for the MISO-IMG-LiDAL and MIMO-LiDAL systems respectively) and $\beta$ is an arbitrary pruning factor (here $\beta \in [1,2..N_{in}]$ [44]). The training process uses the following steps:

i. Provide inputs to the neural network made up of the input samples $X_i$, the initial number of hidden neurons $N_m$ for a given $\beta_i$, the initial weights $(w_{ij})$, connecting the input layer and the hidden layer, the initial weights $(w_{jk})$ connecting the hidden layer and the output layer; and provide the initial biasing input weights $L_j$ and $M_k$.

ii. Calculate the outputs of the hidden neurons associated with the inputs $X_i$, weights $w_{ij}$ and biasing weights $(M_k)$ and apply the result as an input to the neuron activation function resulting in [45], [46]:

$$N_j = S\left(\sum_{i=1}^{N_m} w_{ij}x_i + L_j\right) \qquad j \in [1,..N_m], i \in [1,..N_{in}]$$

(3)

where $S$ is the node activation function, and a sigmoidal function was used, $S(t) = \frac{1}{1+e^{-t}}$.

iii. Calculate the predicted output value $(Y_k)$ depending on the outputs of the hidden neurons $(N_{m_j})$ and their weights $w_{jk}$ with basing weights $(M_k)$, where the output $Y_k$ can be written as [46]:

$$Y_k = \sum_{j=1}^{m} N_j w_{jk} + M_k \qquad k \in [1,..Y] \quad (4)$$

iv. Calculate the error associated with the predicted $(y_k)$ considering the actual number of targets $(A_{Nk})$.

$$e = \sum_{k=1}^{Y} A_k - y_k \quad (5)$$

v. Optimize the weights and biasing. In our approach, we used the Levenberg-Marquardt algorithm (LMA) which is known to be a stable, fast and efficient algorithm with slow error convergence [45], [47]. According to LMA all the network connection weights and biasing weights are updated as defined in [48], [49]:

$$w_{n+1} = w_n - [J_n^T J_n + \mu r I]^{-1} J_n e_n \quad (6)$$

where $\mu r$ is the learning coefficient, $I$ is identity matrix and $J$ is the Jacobian matrix which is calculated as in [48], [49].

vi. Evaluate the error when the new updated weights are used. If the current error is still greater than the required value, update the learning cycle then return to step (ii) with new weight values and learning coefficients.

vii. Update the pruning factor $(\beta_i)$ then return to step (i).

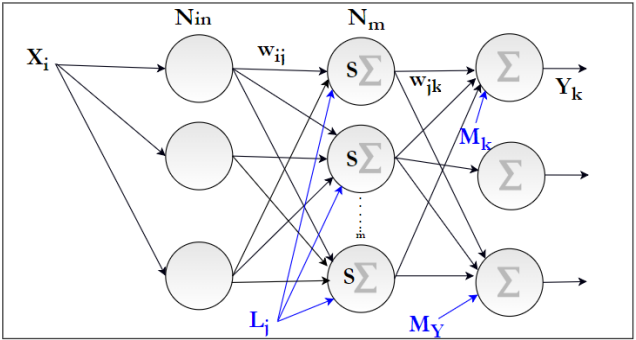

Fig.3: Block diagram of the LiDAL neural network.

We trained the neural network for about 500 learning iterations (epochs) and the learning rate $\mu r$ was 0.05. Larger learning rates can lead to faster convergence. We settled on this smaller value to increase accuracy albeit at the cost of convergence rate. We noticed that beyond 500 learning iterations, the convergence error of the neural network was not significant. We tested a different number of hidden neurons using the pruning approach to find the optimal number of hidden neurons $N_m$ which was 15 and 27 for the MIMO-LiDAL and MISO-IMG-LiDAL systems respectively. The mean square error in the training phase was about $10^{-5}$ validated through 20% of the snapshots.

During the system set-up, the neural network can be trained in situ in the scene, i.e. in the environment where it is to be used. Neural network retraining for any indoor environment can be done through scanning the pulse reflection patterns in the environment in the absence of any targets to determine the obstacles' (furniture) reflection patterns. Targets are then inserted in the environment and the new reflection patterns in the presence of the targets are determined, thus training the ANN.

Fig. 4 presents the proposed block diagram of the ANN receiver for MIMO LiDAL systems. The controller conducts the targets detection and localization process as follows:

1) The control signal activates the transceiver unit of LiDAL to (i) transmit an optical pulse signal from the transmitter $Tx(n)$, and (ii) activates the receiver $Rx(n)$ to collect the reflected signal.

2) The receiver $Rx(n)$ listens to the reflected signal in an observation widow of duration $T$. A trained ANN is activated to process the received signal to detect the targets' presence and their ranges and update the counter as can be seen in Fig. 4.

3) For target localization, the controller finds the $N$ neighbouring LiDAL transmitters. We considered $N=2$ [33], the neighbouring transmitters are $Tx(n+1)$, and $Tx(n+2)$ as can be seen in Fig. 4 in conjunction with the receiver $Rx(n)$. The three trip times (one from $Tx(n)$, and two from neighbouring $Tx(n+1)$, and $Tx(n+2)$) are then used to determine the targets' locations using TOA.

4) Target elimination follows where the targets located in the overlap zones are counted only once in the MIMO LiDAL system (see Fig. 1). Due to position errors, duplicate targets are eliminated then the counter is updated accordingly.

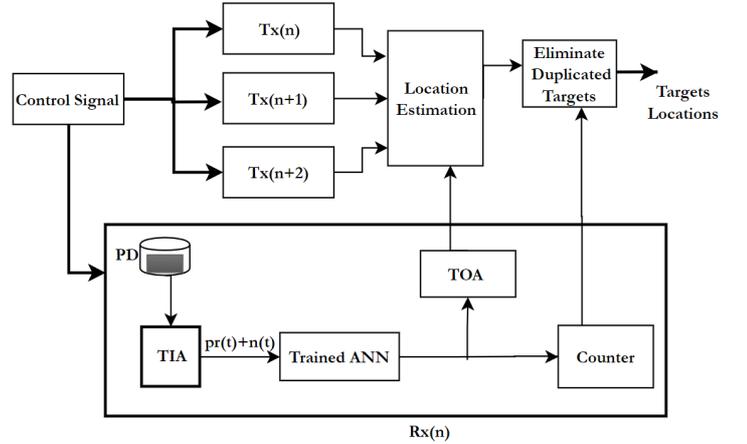

Fig. 4: ANN receiver for MIMO LiDAL system.

Fig. 5 shows the proposed schematic ANN receiver diagram for MISO IMG LiDAL systems. The controller coordinates the targets detection and localization process as follows:

1) The controller activates a transmitter $T_x(n)$ which sends an optical pulse, and also activates the group receiver's pixels $GRP(n)$ to collect the reflected signals.

2) The controller then updates the value of $n$, and if $L_{tx} > n$ step (1) is repeated, where $L_{tx}$ is the number of active transmitter units ($L_{tx}=8$) of the MISO-IMG-LIDAL system [33].

3) A trained ANN is used to process the received reflected signals from each group of receiver pixels to detect and count the targets.

4) Finally, pixel identification is employed to estimate the target location by using DOA method.

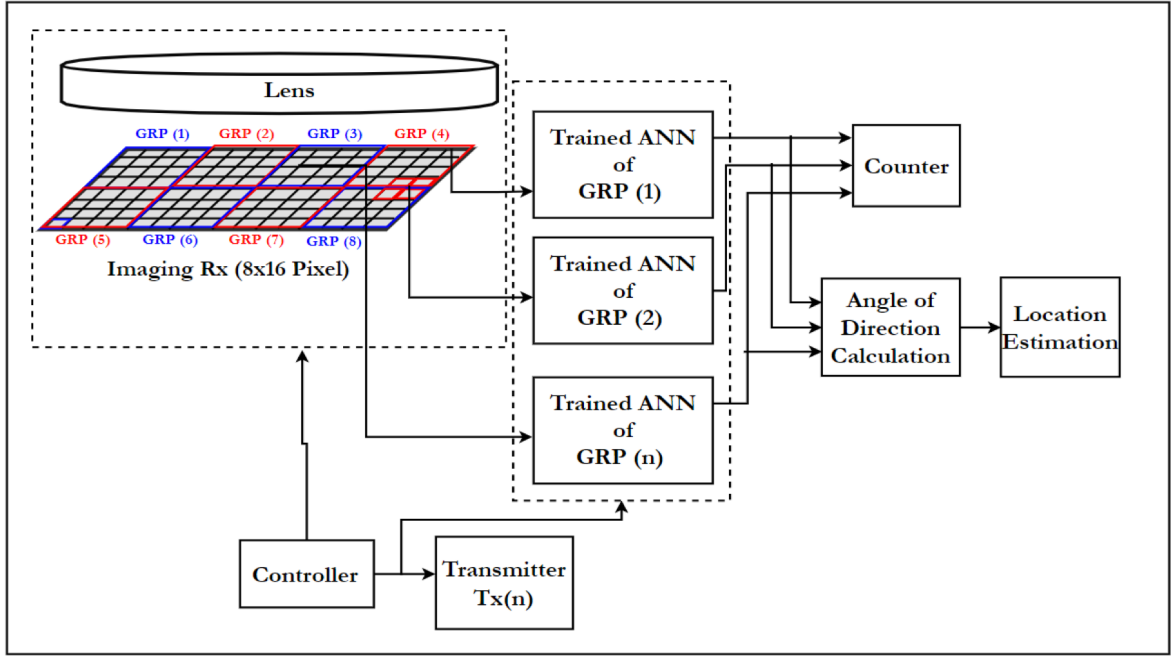

Fig. 5: ANN receiver for MISO IMG LiDAL system

## IV. SIMULATION SETUP AND RESULTS

In this section, we evaluate the performance of the ANN LiDAL systems. We start by assessing the performance of ANN, BSM and CCM in terms of target distinguishing when the environment changes, ie when the locations of furniture change and consider a single target in this case. We then proceed to report results for the ANN-MIMO-LiDAL and ANN-MISO-IMG-LiDAL systems with multiple targets and compare these results to the results reported in [33].

For target mobility, we have considered the directed random walk with obstacle avoiding which was proposed in [33]. In this model, the target walks freely inside the realistic environment in all directions except directions that lead to background obstacles such as furniture [33].

We have considered a realistic office environment which was reported in [33]. The environment consists of a furnished room, with dimensions of 4 m (width) × 8 m (length) × 3 m (height). The reflection factors for the walls and ceiling were 0.8 and 0.3 respectively. The furniture consist of four office desks (1.54 m (width) × 0.76 m (length) × 0.75 m (height)) and one bookshelf (3 m × 0.8 m × 2 m) [33]. The office desks and bookshelf materials were finished-wood with a reflectivity factor of 0.55 and diffuse reflections [26].

The average target of interest (human) dimensions 15 cm × 48 cm × 170 cm (depth × width × height) were considered [33]. The target reflection was considered a Gaussian random variable with a mean of 0.72 and a standard deviation of 0.3 [33].

To evaluate the counting and localization performance of the different LiDAL systems two key metrics are defined: (*i*) The mean absolute percentage error (MAPE) which is used to quantify the counting accuracy, and (*ii*) the distance root means square error (DRMSE) which is used to quantify the localization accuracy [33].

### A. Single Target Distinguishing

We have evaluated the performance of the trained neural network when it distinguishes a single target in a realistic environment, as shown in Fig. 6. The evaluation is conducted in two scenarios, the first scenario included a static realistic environment where the background obstacles (furniture) are fixed over the simulation time with a single nomadic target that moves at a speed of 0.5m/s. The second scenario considered a dynamic realistic environment where the positions of some of the background obstacles (furniture) change over the simulation time in the presence of a nomadic target. A monostatic LiDAL system (collocated transmitter and receiver) was used in the room setup as shown in Fig. 6. In addition, we have considered the pathway model proposed in [33] for target mobility with eight interesting locations ($L_D$=8) in the room in Fig. 45. Five snapshot measurements per second were collected to capture the target movement during the 5 minutes simulation time. The total number of recorded snapshot measurements was 1500.

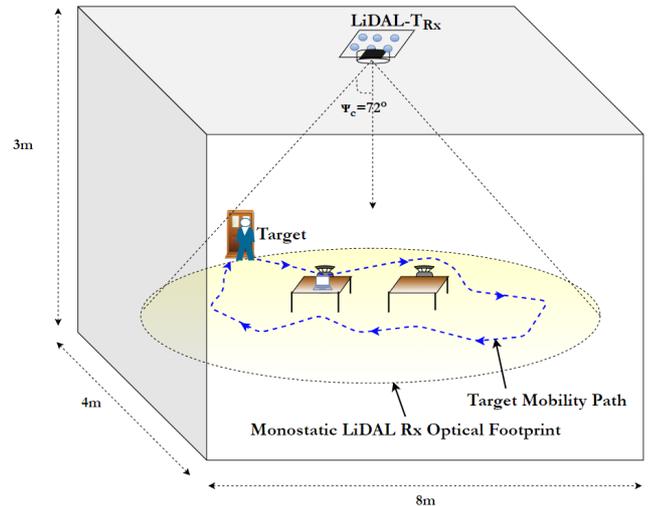

Fig.6: Simulation room setup with monostatic LiDAL.

MAPE was evaluated for the scenario described above and shown in Fig. 6, where the number of targets is one, however, a large number of snapshot measurements were taken as above, while the target moves. The ANN reports target results for each snapshot. The BSM and CCM used two consecutive snapshots. Fig. 7 presents MAPE results, referred to here as the average (over the 1500 snapshot experiment) false distinguishing error for the first scenario, ie the static environment. The ANN was pre-trained and optimized for the room shown in Fig. 6. As can be noted in Fig. 7, the ANN has better performance with 8% error compared to CCM and BSM which have 11% and 19% error respectively. Note that in this experiment, there is a single moving target, the furniture is stationary and the percentage error reflects the ability of the methods to distinguish a moving target from furniture over the large number of snapshots considered.

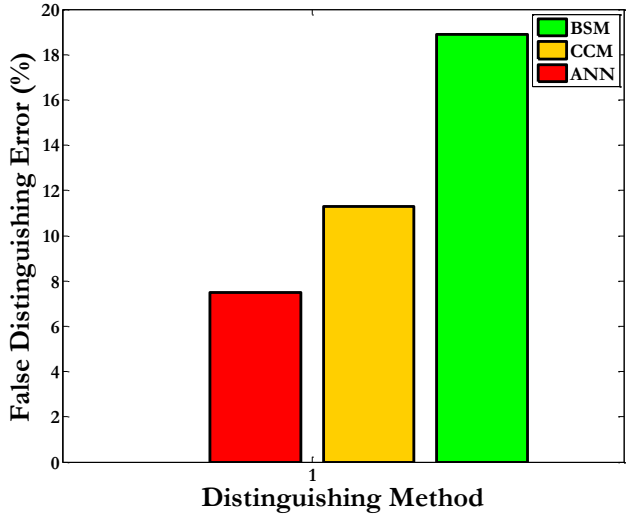

Fig. 7 BSM, CCM and ANN target distinguishing error.

The BSM has the worst performance due to the impact of target presence and movement on the reflections from the background furniture and the particular sensitivity of subtraction to such changes. The ANN and CCM performs better, however, this method fails to distinguish the target only when target-furniture ambiguity occurs. In other words, when the distance between the target and furniture is less than the LiDAL resolution of 0.3m and at the same time, the target remains stationary, (nomadic), for more than 5 snapshots in our experiment. The CCM performance can be improved if the number of processed snapshots is increased to accommodate target mobility behaviour, however, this may slow the target detection process in LiDAL systems.

Fig. 8 shows the average false target distinguishing error percentage for the second scenario, ie a dynamic environment. We simulated the impact of the change in the environment, ie change in furniture configurations as can be seen in Fig. 6 where the furniture positons were changed in each simulation. Note that, the ANN was calibrated and optimized before and after the target presence, but the furniture locations remained fixed throughout the training phase. As can be noted in Fig. 8, the ANN has the best performance up to 40% change in the locations of furniture (tables in this case). If the furniture locations change by a larger percentage, the CCM performs better. This is attributed to the fact that a change in the furniture locations affects the CCM once only, ie when it happens. Beyond that point, the furniture remains static in its new position and the CCM is thus able to track the moving target. The ANN fails as the environment is now significantly different to that over which it was trained. The sensitivity of the BSM is high throughout as explained earlier. The ANN has an error of 35% with 100% change in the environment. This 100% change in our case means that the two tables move from their initial positions at the centre of the room where they are each separated by 0.5m from the centre point of the room, to new locations next to the walls, a 2m movement for the 1.5m × 0.9m table. The BSM and CCM performed better than the ANN at 100% change in the environment, with a maximum error of 27% and 13% for BSM and CCM respectively.

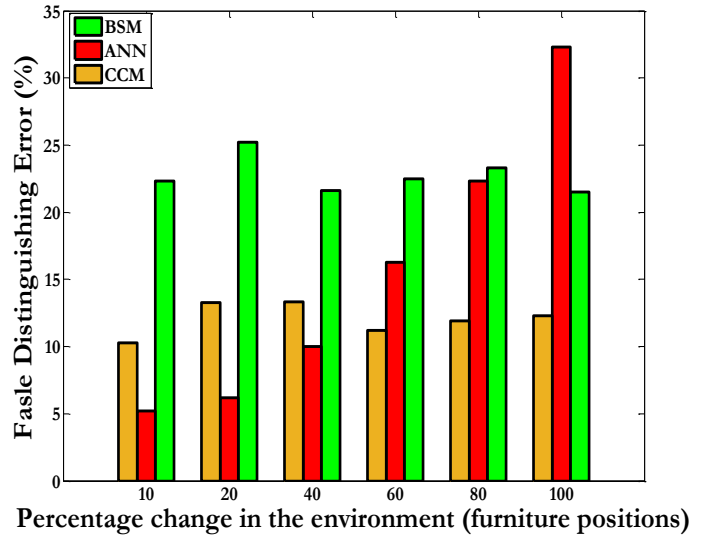

Fig.8: False target distinguishing error in a dynamic environment.

*B. LiDAL Systems with ANN*

We tested the performance of the MIMO-LiDAL and MISO-IMG-LiDAL systems with ANN, BSM and CCM. Table I illustrates the simulation parameters of LiDAL systems. For fair comparison, we have considered the simulation environment and simulation parameters proposed in [33]. The LiDAL systems were evaluated in a scenario which represents '*a challenging localization environment*' reported in [33]. It is worth mentioning that in challenging localization environments, there are multiple moving targets that move continuously such as pedestrians, and there are stationary obstacles (furniture) [33]. The ANN was trained in the environment in the absence of targets and then in their presence. In addition, we used the simulation approach illustrated in Table II to simulate and collect data (snapshot measurements) for the neural network training for both MIMO-LiDAL and MISO-IMG LiDAL systems. We set the number of iterations to $Itr$=250 with 10 snapshots per iteration with $i_{max}$=15.

Fig.9. shows the counting MAPE results for LiDAL systems that include and exclude ANNs. The LiDAL systems that do not employ ANNs, use CCM, the better of the two distinguishing methods. It can clearly be seen in Fig. 9 that the counting MAPE of ANN MIMO-LiDAL is 5% which is significantly lower than the corresponding value, 16%, for MIMO-LiDAL with CCM and the sub-optimum receiver reported in [33]. Furthermore, the performance of MISO IMG-LiDAL with ANN improves, with a maximum counting error of 2%.

Fig. 10 shows the cumulative distribution function of the DRMSE positioning error for the MIMO-LiDAL system and MISO IMG LiDAL with ANN and CCM. As can be noted, the 95% CDF confidence interval is at 0.5m and 0.42m positioning error for MIMO-LiDAL system with CCM and ANN respectively, while the average DRMSE is 0.37m and 0.4m respectively.

It should be observed that overall, the DRMSE values in MISO-IMG-LIDAL are smaller than the corresponding values in MIMO-LiDAL. In the MISO-IMG-LiDAL system, at the 95% confidence interval, Fig. 10, the DRMSE are 0.23m and 0.2m for MISO-IMG-LIDAL with CCM and ANN respectively, whereas the average values of DRMSE are 0.2m and 0.18m for CCM and ANN respectively.

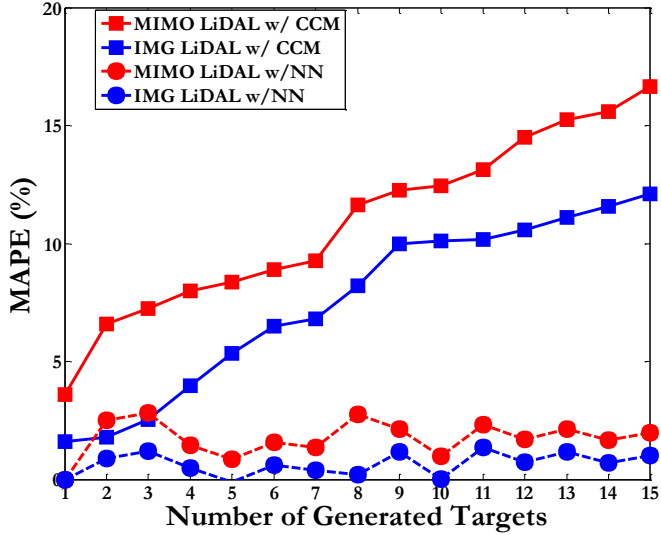

Fig.9: Counting MAPE of LiDAL systems using a trained neural network in a realistic environment.

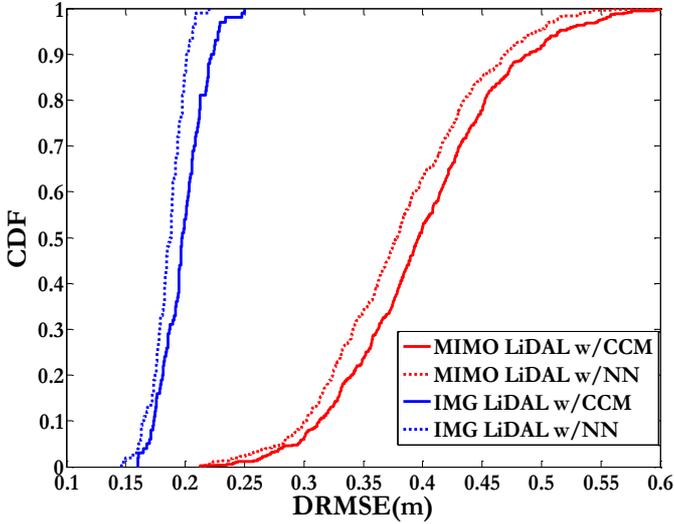

Fig.10: CDF of DRMSE of the MIMO LiDAL and IMG LiDAL systems with CCM and ANN.

## V. CONCLUSIONS

This paper presented new localization systems that employ artificial neural networks with MIMO LiDAL and MISO IMG LiDAL systems for people detection, counting and localization. The results of intelligent LiDAL systems with the trained neural network show that significant improvement in the counting and localization are achieved compared with traditional LiDAL systems with distinguishing methods namely; BSM and CCM.
The best performance for our LiDAL systems was obtained when an ANN with forward backward propagation was used for target detection. The MIMO-LiDAL system with ANN in scenario 2 reduced the counting MAPE to 2% from the 16% associated with the MIMO-LiDAL system. In the MISO-IMG-LiDAL system the use of the ANN reduced the counting MAPE from 12% to approximately 1%. Furthermore, we studied the impact of training the ANN on a given room, and subsequently changing the furniture locations in the room. In a monostatic configuration with a single target, the counting MAPE was below 11% for up to 40% change in the room furniture locations showing high ANN robustness. For furniture location changes beyond 40%, the CCM performs better than ANN as it is able to adapt to the new furniture locations, unlike the ANN which is pre-trained. It is highly likely though that typical changes in room furniture locations will be below 40%, and if above this level, the ANN can include new self-training routines.

TABLE I
SIMULATION PARAMETERS

| Parameters | Configurations |
|---|---|
| Realistic Environment | |
| Length | 8m |
| Width | 4m |
| Height | 3m |
| $\rho$ - ceiling | 0.8 |
| $\rho$ - floor | 0.3 |
| $\rho$ - walls | 0.8 |
| LiDAL Transmitter Units | |
| locations $(x, y, z)$ | (1,1,3), (1,3,3), (1,5,3), (1,7,3) (3,1,3), (3,3,3), (3,5,3), (3,7,3)m |
| Elevation | $90^0$ |
| Azimuth | $0^o$ |
| RGB-LDs in each unit | 9 (3×3) |
| Transmitted optical power per unit | 18 W |
| Transmitted Pulse width $\tau$ | 2ns |
| RGB-LD semi-angle at half power beam width ($\Phi$) | $75^o$ |
| MIMO LiDAL Receiver | |
| Photodetector Area | 20 mm² |
| Receivers locations | Attached with Tx units |
| Photodetector Responsivity | 0.4 A/W |
| Receiver Acceptance Semi-angle | $43.8^o$ |
| CPC Reflective Index ($N$) | 1.7 |
| TIA Noise Current | 2.5 pA/√Hz |
| MISO IMG LiDAL Receiver | |
| Photodetector Area | 2cm² |
| Receiver location $(x, y, z)$ | (2,4,3)m |
| Number of pixels | 128 |
| Pixel's area | 1.56 mm² |
| TIA Pixel Receiver Noise Current | 2.6 pA/√Hz |
| Lens FOV | $72^o$ |
| Time Bin Duration | 0.01 ns |
| Sampling Time $T_{sa}$ | 0.1ns |
| Time Slot Width $T_s$ | 2ns |
| Listening Time $T$ | 1ms |

**TABLE II: SIMULATION FLOW AND DATA COLLECTIONS FOR ANN TRAINING**

| | |
|---|---|
| **Inputs:** | $i_{max} = K$; (Maximum number of targets) |
| | $j_{max} = Itr$; (Number of iterations) |
| | $\varrho(k,i)$ is target $k$ reflection factor when an environment with $i$ targets is considered |
| 1. | **for** $i = 1: i_{max}$; |
| 2. | **for** $j = 1: j_{max}$; |
| 3. | Generate a random location(s) $l(k,j)$ and $\varrho(k,i)$ for target(s) $k \in [1,..i]$ |
| 4. | Generate additive white Gaussian noise $n_j(t)$ |
| 5. | Apply LiDAL system detection algorithm |
| 6. | $j == j_{max}$ |
| 7. | **end for** |
| 8. | Calculate MAPE |
| 9. | Calculate DRMSE |
| 10. | save MAPE and DRMSE at given value of $i$ |
| 11. | $i == i_{max}$ |
| 12. | **end for** |

ACKNOWLEDGEMENTS

Aubida A. Al-Hameed would like to thank the Higher Committee for Education Development in Iraq (HCED) and the University of Mosul for financial support and funding his PhD scholarship.

This work was supported by the Engineering and Physical Sciences Research Council (ESPRC), INTERNET (EP/H040536/1), and STAR (EP/K016873/1) projects. All data are provided in full in the results section of this paper.

**Aubida A. Al-Hameed** received a B.Sc. in electronic and electrical engineering from the University of Mosul, Iraq, in 2010 and an M.Sc. degree in communication systems from the University of Mosul, Iraq, in 2013. He is a Higher Committee for Education Developments in Iraq (HCED) Scholar and is currently working towards a Ph.D. degree in the school of Electronic and Electrical Engineering, University of Leeds, Leeds, UK.

**Safawn Hafeedh Younus** received the B.Sc. degree in electronic and electrical engineering, and the M.Sc. degree (Hons.) in communication systems from the University of Mosul, Iraq, in 2008 and 2010, respectively. He is currently pursuing the Ph.D. degree with the School of Electronic and Electrical Engineering, University of Leeds, Leeds, U.K. He is a 'Ministry of Higher Education and Scientific Research (MOHESR) of Iraq' Scholar. Prior to his Ph.D. study, he was with the Ministry of Communication, Iraq, as a Maintenance and Support Engineer of the NGN Local Network, from 2010 to 2012. He also worked as a Lecturer with the Communication Department, College of Electronics, University of Mosul, from 2012 to 2014. His research interests include performance enhancement techniques for visible light communication systems, visible light communication system design, and indoor visible light communication networking.

**Ahmed Taha Hussein** received the B.Sc. degree (Hons.) in electronic and electrical engineering and the M.Sc. degree (Hons.) in communication systems from the University of Mosul, Iraq, in 2006 and 2011, respectively, and the Ph.D. degree in visible light communication systems from the University of Leeds, Leeds, U.K., in 2017. Prior to his Ph.D. study, he worked as a Communication Instructor with the Electronic and Electrical Engineering Department, College of Engineering, University of Mosul, from 2006 to 2009. He also worked as a Lecturer at the Electronic and Electrical Engineering Department, College of Engineering, University of Mosul, from 2011 to 2012. He published widely in the top IEEE communications conferences and journals. He has received the Carter award, University of Leeds, for the best Journal. His research interests include performance enhancement techniques for visible light communication systems, visible light communication.

**Mohammed Thamer Alresheed** received the B.Sc. degree (Hons.) in electrical engineering from King Saud University, Riyadh, Saudi Arabia, in 2006, and the M.Sc. degree (Hons.) in communication engineering, and the Ph.D. degree in electronic and electrical engineering, both from Leeds University, Leeds, U.K., in 2009 and 2014, respectively. He is currently an Assistant Professor with the Department of Electrical Engineering, King Saud University. His research interests include adaptive techniques for optical wireless (OW), OW systems design, indoor OW networking, and visible light communications.

**Prof. Jaafar M. H. Elmirghani** is the Director of the Institute of Integrated Information Systems within the School of Electronic and Electrical Engineering, University of Leeds, UK. He joined Leeds in 2007 and prior to that (2000–2007) as chair in optical communications at the University of Wales Swansea he founded, developed and directed the Institute of Advanced Telecommunications and the Technium Digital (TD), a technology incubator/spin-off hub. He has provided outstanding leadership in a number of large research projects at the IAT and TD. He received the Ph.D. in the synchronization of optical systems and optical receiver design from the University of Huddersfield UK in 1994 and the DSc in Communication Systems and Networks from University of Leeds, UK, in 2014. He has co-authored Photonic switching Technology: Systems and Networks, (Wiley) and has published over 500 papers. He has research interests in optical systems and networks. Prof. Elmirghani is Fellow of the IET, Fellow of the Institute of Physics and Senior Member of IEEE. He was Chairman of IEEE Comsoc Transmission Access and Optical Systems technical committee and was Chairman of IEEE Comsoc Signal Processing and Communications Electronics technical committee, and an editor of IEEE Communications Magazine. He was founding Chair of the Advanced Signal Processing for Communication Symposium which started at IEEE GLOBECOM'99 and has continued since at every ICC and GLOBECOM. Prof. Elmirghani was also founding Chair of the first IEEE ICC/GLOBECOM optical symposium at GLOBECOM'00, the Future Photonic Network Technologies, Architectures and Protocols Symposium. He chaired this Symposium, which continues to date under different names. He was the founding chair of the first Green Track at ICC/GLOBECOM at GLOBECOM 2011, and is Chair of the IEEE Green ICT committee within the IEEE Technical Activities Board (TAB) Future Directions Committee (FDC), a pan IEEE Societies committee responsible for Green ICT activities across IEEE, 2012-present. He is and has been on the technical program committee of 34 IEEE ICC/GLOBECOM conferences between 1995 and 2015 including 15 times as Symposium Chair. He received the IEEE Communications Society Hal Sobol award, the IEEE Comsoc Chapter Achievement award for excellence in chapter activities (both in 2005), the University of Wales Swansea Outstanding Research Achievement Award, 2006, the IEEE Communications Society Signal Processing and Communication Electronics outstanding service award, 2009, a best paper award at IEEE ICC'2013, the IEEE Comsoc Transmission Access and Optical Systems outstanding Service award 2015 in recognition of "Leadership and Contributions to the Area of Green Communications" and received the GreenTouch 1000x award in 2015 for "pioneering research contributions to the field of energy efficiency in telecommunications". He is currently an editor of: IET Optoelectronics, Journal of Optical Communications, IEEE Communications Surveys and Tutorials and IEEE Journal on Selected Areas in Communications series on Green Communications and Networking. He is Co-Chair of the GreenTouch Wired, Core and Access Networks Working Group, an adviser to the Commonwealth Scholarship Commission, member of the Royal Society International Joint Projects Panel and member of the Engineering and Physical Sciences Research Council (EPSRC) College. He has been awarded in excess of £22 million in grants to date from EPSRC, the EU and industry and has held prestigious fellowships funded by the Royal Society and by BT. He is an IEEE Comsoc Distinguished Lecturer 2013-2016.